# ON THE GINZBURG-LANDAU ANALYSIS OF THE LOWER CRITICAL FIELD $H_{c1}$ IN $MgB_2$

**İ.N.Askerzade[1,2], A. Gencer[1,+] N.Güçlü[1] and A.Kılıç[1]**

[1]**Ankara University, Faculty of Sciences, Department of Physics,**

**06100-Tandoğan-Ankara/Turkey**

[2] **Institute of Physics, Azerbaijan Academy of Sciences, Baku, 370143-Azerbaijan**

## Abstract

Temperature dependence of the lower critical field $H_{c1}$ (T) for the superconducting magnesium diboride, $MgB_2$, is studied in the vicinity of $T_c$ by using a two-band Ginzburg-Landau (G-L) theory. Theoretically calculated $Hc_1$ (T) near $T_c$ exhibits a negative curvature. The results are shown to be in a good agreement with the experimental data. In addition, two-band G-L theory calculations give a temperature dependence of the Ginzburg-Landau parameter $\kappa(T) = \frac{\lambda(T)}{\xi(T)}$, which varies little with temperature as in a similar manner of the microscopic single-band BSC theory.

**Keywords:** Superconductivity, Macroscopic Theory, The Lower Critical Field $H_{c1}$, Two-Band Superconductivity and Negative Curvature.

[+] Corresponding Author: gencer@science.ankara.edu.tr, Tel : 90-312-212 67 20/Ext. 1161.

## I. INTRODUCTION

The recent discovery of superconductivity in an intermetallic compound of magnesium diboride, $MgB_2$, [1] is interesting for some reasons. First, $MgB_2$ holds the record of the highest critical temperature at around 40 K among the simple binary compounds of superconductors. The second, the material shows a pronounced isotope effect of boron, which suggests that $MgB_2$ is a phonon mediated superconductor [2]. In addition, the compound has been found particularly attractive because it is generally thought that $MgB_2$ has a potential to replace conventional superconducting materials in the electronic applications in terms of large critical current densities [3] and transparency between grain boundaries to the current flow [4]. $MgB_2$ is of a type II superconductor with the estimated values of a coherence length $\xi(0K)$, penetration depth $\lambda(0K)$ and G-L parameter, $\kappa(0)$ being 5.2 nm, 125-140 nm, $\kappa(0)=26$, respectively [5,6].

Microscopic mechanisms of superconductivity in $MgB_2$ and the discussions of its superconducting properties implicitly involve the employment of Isotropic Single Band (ISB) model [7]. However, band structure calculations suggest that the origin of the superconductivity is based on the hole-like boron $\sigma$ band [8] and $\pi$ band [9]. Hence, two-band Eliashberg theory [10] is generally assumed as complementary to the ISB model. Recently two-band Eliashberg model was applied for determining the behaviour of $H_{c2}$ in $MgB_2$ [11].

The temperature dependence of the lower critical magnetic field $H_{c1}$ exhibits a negative curvature near critical temperature $T_c$ as reported experimentally in Refs. [12-14]. For a review, see also the recent work of Ref. [15]. Although there has been some extensive theoretical works on $MgB_2$, to our best of knowledge, there is no systematic model to explain the temperature dependence of $H_{c1}$, observed in $MgB_2$. In a previous communication, we presented the temperature dependence of $H_{c2}$ (T) by using the two-band G-L theory [16]. In this paper, we show that the presence of two order parameters for two bands in the theory yields a negative curvature in the vicinity of the critical temperature for $MgB_2$ superconductor. The calculated temperature dependence of the lower critical field $H_{c1}(T)$ is in a good agreement with some experimental data.

## II- THEORY

We write the G-L free energy functional for two-band superconductors with two coupled superconducting order parameters of the following form similarly as in [16,17],

$$F_{SC}[\Psi_1, \Psi_2] = \int d^3r \left( F_1 + F_{12} + F_2 + H^2/8\pi \right) \tag{1}$$

with

$$F_i = \frac{\hbar^2}{4m_i} \left| \left( \nabla - \frac{2\pi i \vec{A}}{\Phi_0} \right) \Psi_i \right|^2 + \alpha_i(T) \Psi_i^2 + \beta_i \Psi_i^4 / 2 \tag{2}$$

$$F_{12} = \varepsilon(\Psi_1^* \Psi_2 + c.c.) + \varepsilon_1 \left\{ \left( \nabla + \frac{2\pi i \vec{A}}{\Phi_0} \right) \Psi_1^* \left( \nabla - \frac{2\pi i \vec{A}}{\Phi_0} \right) \Psi_2 + c.c. \right\} \tag{3}$$

Here $m_i$ denotes the the effective mass of the carriers belonging to the band i (i=1,2). $F_i$ is the free energy of the separate bands. $F_{12}$ is the interaction energy term between the bands. The coefficient $\alpha$ depends linearly on the temperature T as $\alpha_i = \gamma_i (T - T_{ci})$, while the coefficient $\beta_i$ is independent of temperature. Here, $\gamma_i$ is the proportionality constant. The quantities $\varepsilon$ and $\varepsilon_1$ describe the interband mixing of two order parameters and their gradients, respectively. H is the external magnetic field , $\Phi_0$ is the magnetic flux quantum and $\vec{A}$ is the vector potential. In Eqns. (2) and (3), it is assumed that the order parameter $|\Psi_i|^2$ are slowly varying functions in space.

For temperatures near $T_c$ and magnetic fields slightly larger than $H_{c1}$, the influence of the field on the modulus of the order parameters $\Psi_1$ and $\Psi_2$ can be neglected and we assume $|\Psi_1|$=const , $|\Psi_2|$ =const . Then the representing wave function $\Psi_i(\vec{r})$ can be written as $\Psi_i(\vec{r}) = |\ \Psi_i\ | \exp(i\phi_j(\vec{r}))$. Here, $\phi_j(r)$ are the phases of the order parameters and the G-L free energy functional (1), can be rewritten as

$$\mathbf{F_{SC}}[\phi_1,\phi_2] = \int d^3r \begin{pmatrix} \frac{\hbar^2}{8m_1} n_1(T)(\frac{d\phi_1}{dr} - \frac{2\pi}{\Phi_0}\vec{A})^2 + \frac{\hbar^2}{8m_2} n_2(T)(\frac{d\phi_2}{dr} - \frac{2\pi}{\Phi_0}\vec{A})^2 + \\ + \varepsilon(n_1(T)n_2(T))^{1/2}\cos(\phi_1 - \phi_2) + \\ + \varepsilon_1(n_1(T)n_2(T))^{1/2}\cos(\phi_1 - \phi_2)(\frac{d\phi_1}{dr} - \frac{2\pi}{\Phi_0}\vec{A})(\frac{d\phi_2}{dr} - \frac{2\pi}{\Phi_0}\vec{A}) + H^2/8\pi \end{pmatrix} \qquad (4)$$

Where, $n_1(T) = 2\ |\Psi_1|^2$ and $n_2(T) = 2\ |\Psi_2|^2$ are the densities of superconducting electrons for the corresponding bands, respectively. The temperature dependencies of $n_1$ (T), $n_2$ (T) and ($\phi_1 - \phi_2$) are defined by the equilibrium value of order parameters $|\Psi_1|$ and $|\Psi_2|$, which satisfy the G-L equations (see Eqn. A2, A3).

The equations determining the equilibrium values of magnetic field and phases of the order-parameters would be obtained by the minimising the free energy functional (4) with respect of the vector potential $\vec{A}$ and the phases $\phi_1$, $\phi_2$. The equation for the vector potential takes the form

$$\begin{gathered} \frac{\nabla x \nabla x \vec{A}}{4\pi} = \frac{2\pi}{\Phi_0}\{\ \frac{\hbar^2}{4m_1} n_1(T)(\frac{d\phi_1}{dr} - \frac{2\pi}{\Phi_0}\vec{A}) + \frac{\hbar^2}{4m_2} n_2(T)(\frac{d\phi_2}{dr} - \frac{2\pi}{\Phi_0}\vec{A}) + \\ \varepsilon_1(n_1(T)n_2(T))^{1/2}\cos(\phi_1 - \phi_2)[(\frac{d\phi_1}{dr} - \frac{2\pi}{\Phi_0}\vec{A}) + (\frac{d\phi_2}{dr} - \frac{2\pi}{\Phi_0}\vec{A})]\ \} \end{gathered} \qquad (5)$$

By using the relevant Maxwell equation, $\nabla x \vec{H} = \frac{4\pi}{c} J$ (for the magnetostatic case), Eq. (5) yields to the London equation ( by taking into account the equilibrium value of the differences of phases $\phi_1 - \phi_2$, see A7 and A8) of the form

$$\lambda^2 \frac{d^2H}{dr^2} - H = 0\ , \qquad (6)$$

where $\lambda$ is the London penetration depth of the following form

$$\lambda^{-2}(T) = \frac{4\pi e^2}{c^2}\left(\frac{n_1(T)}{m_1} + 2\varepsilon_1(n_1(T)n_2(T))^{1/2} + \frac{n_2(T)}{m_2}\right) \qquad (7)$$

It is well known that, the lower critical field $H_{c1}$ can be obtained as in Ref. [18]

$$H_{c1} = \frac{\Phi_0}{4\pi\lambda^2(T)} \ln \kappa(T) . \tag{8}$$

We then introduce a dimensionless lower critical fi eld $h_{c1} = \frac{H_{c1}(T)}{H_{c1}(0)}$ with

$H_{c1}(0) = \frac{\Phi_0 e^2 T_c}{c^2}\left(\frac{\gamma_1}{\beta_1 m_1} + \frac{\gamma_2}{\beta_2 m_2}\right)$. We rewrite the normalised lower critical field as:

$$h_{c1} = B(T) \ln \kappa(T) , \tag{9}$$

where

$$B(T) = -\frac{2}{x + D^{-1}}\left(\varepsilon^2 + x(\tau - \tau_{c1})^2 + 2x\eta\varepsilon^2(\tau - \tau_{c1})\right)\left\{\frac{\theta^2 + (2 - \tau_{c1} - \tau_{c2})\theta}{\varepsilon^2 D(\tau - \tau_{c2}) + (\tau - \tau_{c1})^3}\right\} \tag{10}$$

$$D = \frac{\beta_1\gamma_2^2}{\beta_2\gamma_1^2} , \tau_{c1,c2} = \frac{T_{c1,c2}}{T_c} , \tau = \frac{T}{T_c} , x = \frac{\gamma_1 m_1}{\gamma_2 m_2}, \theta = \tau - 1 , \eta = \frac{\varepsilon_1 T_c \gamma_2 m_2}{\varepsilon \hbar^2} \tag{11}$$

Temperature dependence of the normalised G-L parameter $\kappa(T)$ in two-band SC is given as

$$\frac{\kappa(T)}{\kappa(0)} = \left(\frac{h_{c2}(T)}{B(T)}\right)^{1/2} , \tag{12}$$

Here, the upper critical field $h_{c2}$ of two-band superconductors was calculated as in [16] and can be given as

$$h_{c2} = a^{-1}\left(-\theta - c_0 + (A\theta^2 + B\theta + c_0)^{1/2}\right) \tag{13}$$

The parameters in last equation are the same as in [16].

## III- RESULTS AND DISCUSSIONS

The experimental data for the lower critical field $h_{c1}$ of $MgB_2$ can be fitted well with high accuracy by the simple expression of the form

$$h_{c1} = h_{c1}^*(-\theta)^{1-\alpha} . \tag{14}$$

In figure 1, we plot the experimental data of Li et al. [12] with open circle symbol and Sharoni et al. [14] with filled circle symbol for $H_1(T)$ in $MgB_2$ versus the reduced temperature $T/T_c$ with $H_{c1}(0)$ = 290 Oe and 450 Oe respectively. We also show the theoretical fits to the data by using Eqns.(9) and (14). The best fit to the experimental data was obtained by using Eqn.(14) (shown by solid lines) with fitting parameters $h^*_{c1} = 0.9$, $\alpha = 0.23$. As shown in [16], the upper critical field can be fitted by the similar formula with $\theta^{1+\alpha}$ , $\alpha = 0.23$ In Fig.1, the dotted line displays the two-band G-L fitting by using Eqn.(9) with D=1.35. The temperature dependence of the two-band G-L parameter $\frac{\kappa(T)}{\kappa(0)}$ is shown in Fig. 2. In the calculations, G-L parameters for the fitting $h_{c2}$ are the same as in [16]

$$A = 0.59, B = -0.11, c_0 = 0.20, x = 3, \mu_0 \tilde{H}_{c2}(0) = 15.55\ T, T_{c1} = 20\ K, T_{c2} = 10\ K.$$

Calculations show that the function $\frac{\kappa(T)}{\kappa(0)}$ has weak temperature dependence as in a similar way of the single-band BCS theory [18] implying that the temperature dependence is determined mainly by the temperature dependence of the penetration depth. On the other hand, as followed from the equation (7), $h_{c1}$ is determined dominantly by the small mass of the carriers in contrast to the upper critical field, $h_{c2}$. The term in Eqn. (7) for the intergradient interaction between the bands may also contribute to the temperature dependence of the penetration depth. Inclusion of the term with $\varepsilon_1$ enhance the negative curvature near critical temperature (dashed line) in similar manner for upper critical field. Thus, in two-band G-L theory, the lower critical magnetic field exhibits negative curvature in contrast to ISB GL theory. Note that the single band G-L theory would only give a linear temperature dependence for $H_{c1}$ [18] only at temperatures close to $T_c$.

As shown by Eqn. (A4) , the critical temperature of around $T_c$ =40 K can be obtained from $T_{c1}$=20 K and $T_{c2}$=10 K approximately with the interaction parameter $\varepsilon^2 = 3/8$. As followed from the eq. (11) lower critical field $h_{c1}$ is governed by the parameter $\varepsilon_1$, $\varepsilon$. Our fitting corresponds to $\eta$ = - 0.16. We also take into account that the average Fermi velocity over hole tubular band is $v_{F1}$=4.5 $10^7$ cm/s, while for the remaining last Fermi surface sheets we can choose $v_{F2}$ =11. $10^7$ cm/s .

Note that the theoretical results of two band superconductivity is in a good agreement with the experimental works of [12] and [14]. However, there are some experimental works that are not well fitted to our results at high temperatures [19] and at low temperatures [13]. There may be several reasons for the inconsistency. First, Joshi et al. [19] has presented experimental measurements of $H_{c1}$ well below $T_c$ and found a linear dependence on temperature. It is generally accepted that $H_{c1}$ becomes vanishigly small near $T_c$ and its measurement is almost impossible at temperatures in the vicinity of Tc. We also note that our theoretical results for hc2 [16] are in agreement with their experimental data of Ref. [19]. In addition, the measurement of $H_{c2}(T)$ in Ref. [13] also shows positive curvature near Tc. Their data for $H_{c1}(T)$ are in a reasonable agreement with the data of Sharoni et al. [14] at temperatures between 20 and 40 K, while their data differ at temperatures below 20 K. In our opinion, the discrepancy may result from the sample specifications, since optimisation of the sample preparation is still needed to make high quality samples. Two-band G-L theory gives better temperature dependence than single-band G-L theory. This may support the proposals that $MgB_2$ is the first two-band superconductor [8,9,11,20]

In summary, we have shown that the negative curvature of the lower critical field can be explained to a reasonable extent by seeking analytical solutions to the two-band G-L theory within an approximation. An exact fitting between experiement and the theory would necessitate exact solutions to the theory. However, the theory itself is phonemologic and it can explain the temperature dependence to some extent only at temperatures close to $T_c$. Further work is therefore needed to elucidate the nature of superconductivity and the interesting physical properties observed in $MgB_2$ at both microscopic level and the macroscopic level.

## IV- ACKNOWLEGMENTS

One of us (I.N.A) is grateful to Professor S.Atag, Head of Physics Deparment at Ankara University for their hospitality during his stay. This work has been financially supported in part by Ankara University Research Fund under a contract number 2000-07-05-001.

## APPENDIX

By minimization of the free energy of Eqn.(1)

$$\delta F / \delta \Psi_1^* = 0 \quad , \quad \delta F / \delta \Psi_2^* = 0 \tag{A1}$$

we obtain the usual basic equation for the description of the two-band superconductivity. For simplicity, we assume a vector potential $\vec{\mathrm{A}} = (0, \mathrm{Hx}, 0)$ in one dimension, then linearisation of Eqn.(4) yields

$$\frac{\hbar^2}{4m_1}(\frac{d^2}{dx^2} - \frac{x^2}{l_s^2})\Psi_1 + \alpha_1(T)\Psi_1 + \varepsilon\Psi_2 + \varepsilon_1(\frac{d^2}{dx^2} - \frac{x^2}{l_s^2})\Psi_2 + \beta_1\Psi_1^3 = 0 \tag{A2}$$

$$\frac{\hbar^2}{4m_2}(\frac{d^2}{dx^2} - \frac{x^2}{l_s^2})\Psi_2 + \alpha_2(T)\Psi_2 + \varepsilon\Psi_1 + \varepsilon_1(\frac{d^2}{dx^2} - \frac{x^2}{l_s^2})\Psi_1 + \beta_2\Psi_2^3 = 0\,. \tag{A3}$$

Where $l_s^2 = (\hbar c/2eH)$ is the so called magnetic length. The coefficients $\varepsilon_1$ and $\varepsilon$ are determined by the microscopic nature of the interaction between different band electrons. Their signs can be taken arbitrarily. If the interband interaction is ignored, the Eqns. (A2) and (A3) are decoupled into two ordinary G-L equations with two different critical temperatures. In general, regardsless of the sign of $\varepsilon$, the superconducting phase transition occurs at a well defined temperature exceeding both $T_{c1}$ and $T_{c2}$ , which can determined by the equation

$$\alpha_1(T_c)\alpha_2(T_c) = \varepsilon^2\,. \tag{A4}$$

In the absence of any magnetic field, the equilibrium value of the order parameters are given as

$$\Psi_1^2 = -\frac{\varepsilon^2\left(\alpha_1(T)\alpha_2(T) - \varepsilon^2\right)}{\varepsilon^2\beta_1\alpha_2(T) + \beta_2\alpha_1^3(T)} \tag{A5}$$

$$\Psi_2^2 = -\frac{\alpha_1^2(T)\ \left(\alpha_1(T)\alpha_2(T) - \varepsilon^2\right)}{\varepsilon^2\beta_1\alpha_2(T) + \beta_2\alpha_1^3(T)} \tag{A6}$$

Phases in the equilibrium state satisfy the following condi tions

$$\cos(\phi_1 - \phi_2) = 1 \text{ or } \phi_1 - \phi_2 = 2\pi n \quad , \quad \text{if } \varepsilon < 0 \tag{A7}$$

$$\cos(\phi_1 - \phi_2) = -1 \text{ or } \phi_1 - \phi_2 = (2n+1)\pi\,, \quad \text{if } \varepsilon > 0. \tag{A8}$$

## REFERENCES

[1] Nagamatsu, J, Nakagawa N., Muranaka T., Zenitani Y., and Akimitsu J. 2001 Nature (London) **410** 63

[2] Bud'ko S., Lapertot G., Petrovic C., Gunningham C.E., Anderson N., Canfield P.C. 2001 Phys.Rev.Lett. **86** 1877

[3] Glowacki B.A. Majoros M., Vickers M., Evetts J.E., Shi Y., McDougall I. 2001 Supercond. Sci.and Technol. **14** 193

[4] Gencer A. 2001 Supercond. Sci.and Technol. accepted for publication in December.

[5 ]X.H. Chen, Y.Y. Xue, R.L.Meng, C.W.Chu 2001 Phys. Rev. B **64** 172501

[6] Finnemore D.K., Osbenson J.E., Budko S.L., Lapertot G., Canfield P.C. 2001 Phys.Rev.Lett. **86** 2420

[7] Carbotte J.P.1990 Rev.Mod.Phys. **62** 1027

[8]Kong Y., Dolgov O.V., Jepsen O., Andersen O.K. 2001 Phys.Rev. B **64**, 020501(R)

[9]Kotrus J., Mazin I.I., Belashenko K.D., Antipov V. P. Boyer L.L. 2001 Phys.Rev.L ett. **86**, 4656

[10] Shulga S.V.,Drechsler S.-L., Muller K.-H, Fuchs G., Winzer K., Heinecke M.and Krug K. 1998 Phys.Rev.Lett**.** **80** 1730

[11] Shulga S.V., Drechsler S.-L., Eschrig H., Rosner H., and Pickett W. , condmat/0103154

[12] Li S.L., Wen H.H., Zhao Z.W., Ni Y.M, Ren Z.A., Che G.C., Yang H.P., Lin Z.Y. and Zhao Z.X. 2001 Phys.Rev.B **64** 094522

[13] Takano Y., Takeya H., Fujii H., Kumakura H., Hatano T., Togano K., Kito H. , Ihara H. 2001 Applied Physics Letters, **78**, 2814

[14] Sharoni A., Felner I.. Millo O. 2001 Phys.Rev.B**63**,220508(R)

[15] Buzea C., Yamashita T.,2001 Supercond.Sci. and Technol. **14**, R115

[16] Askerzade I.N., Gencer A., Güçlü N., 2001, Supercond.Sci. and Technol . november, submitted cond-mat/0112210

[17] Doh H.,Sigrist M., Chao B.K., and Sung-IkLee 1999 Phys.Rev.Lett**.** **85** 5350

[18] Abrikosov A.A. 1988 Fundamentals of the theory of metals, North–Holland,Amsterdam

[19] Joshi A.G., Pillai C.G.S., Raj P., Malik S.K. 2001 Solid state Communication **118,** 445

[20] Schmidt H., Zasadzinski J.F., Gray K.E., Hinks D.G. 2001 cond-mat/0112144

**Figure Captions**

**Figure 1.** Temperature dependence of the upper critical field $H_{c1}(T)$. Open circle symbols show the experimental data of Li et al. in Ref. [12], while open square symbols show the experimental data of Sharoni et al. in Ref. [14]. The solid line shows the calculations from Eq.(14). Filled circle symbols with dotted line represent the two band G-L theory ($\varepsilon_1 = 0$). Triangle symbols with dashed lines are the two-band theoretical calculations of G-L theory ($\varepsilon_1 = 0.0976$) involving interaction between bands.

**Figure 2.** Temperature dependence of Ginzburg-Landau parameters $\kappa$. The solid line represents calculations from the theory with $\varepsilon_1 = 0$ and the dashed line is for $\varepsilon_1 = 0.0976$.

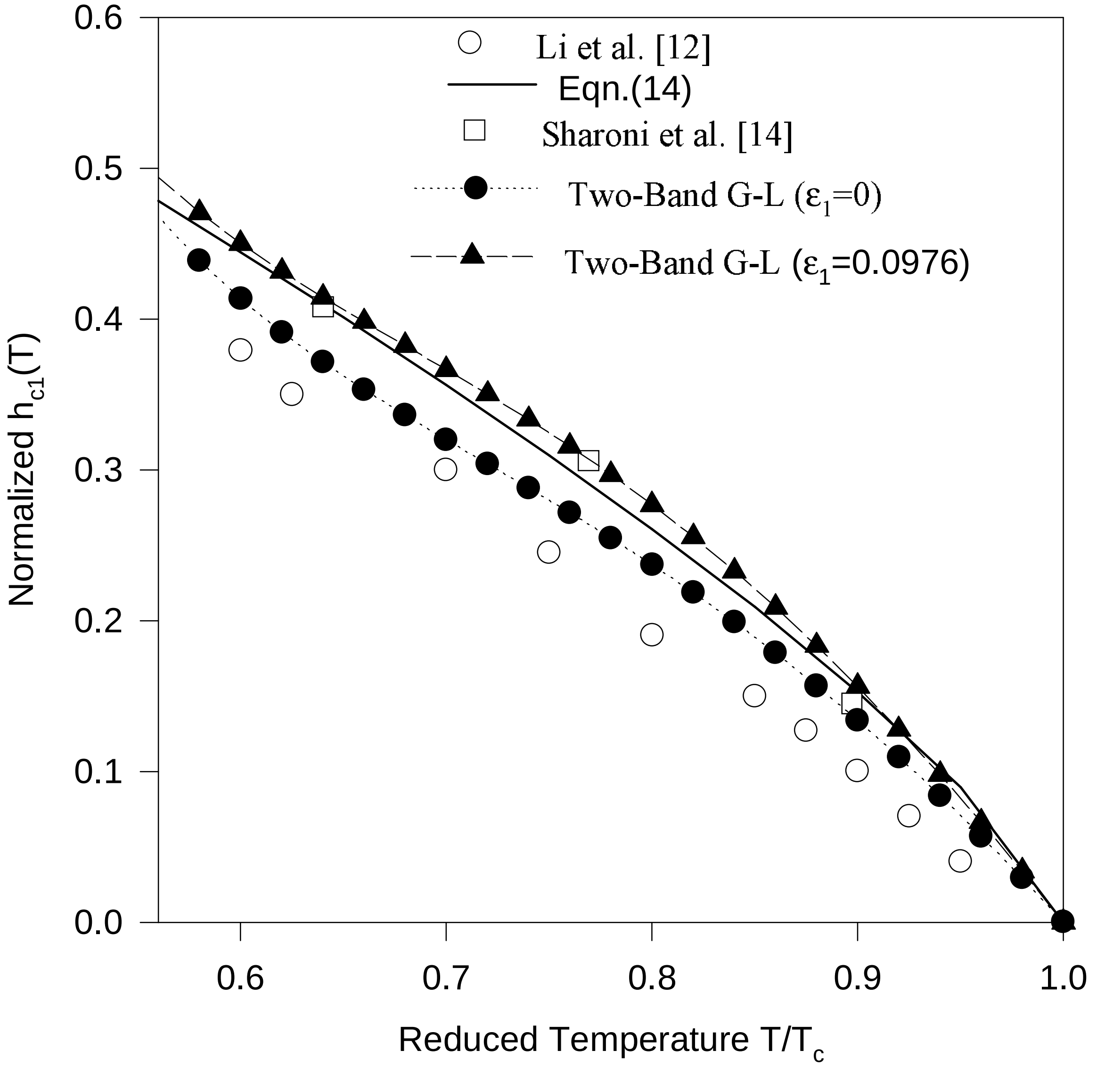

Figure 1

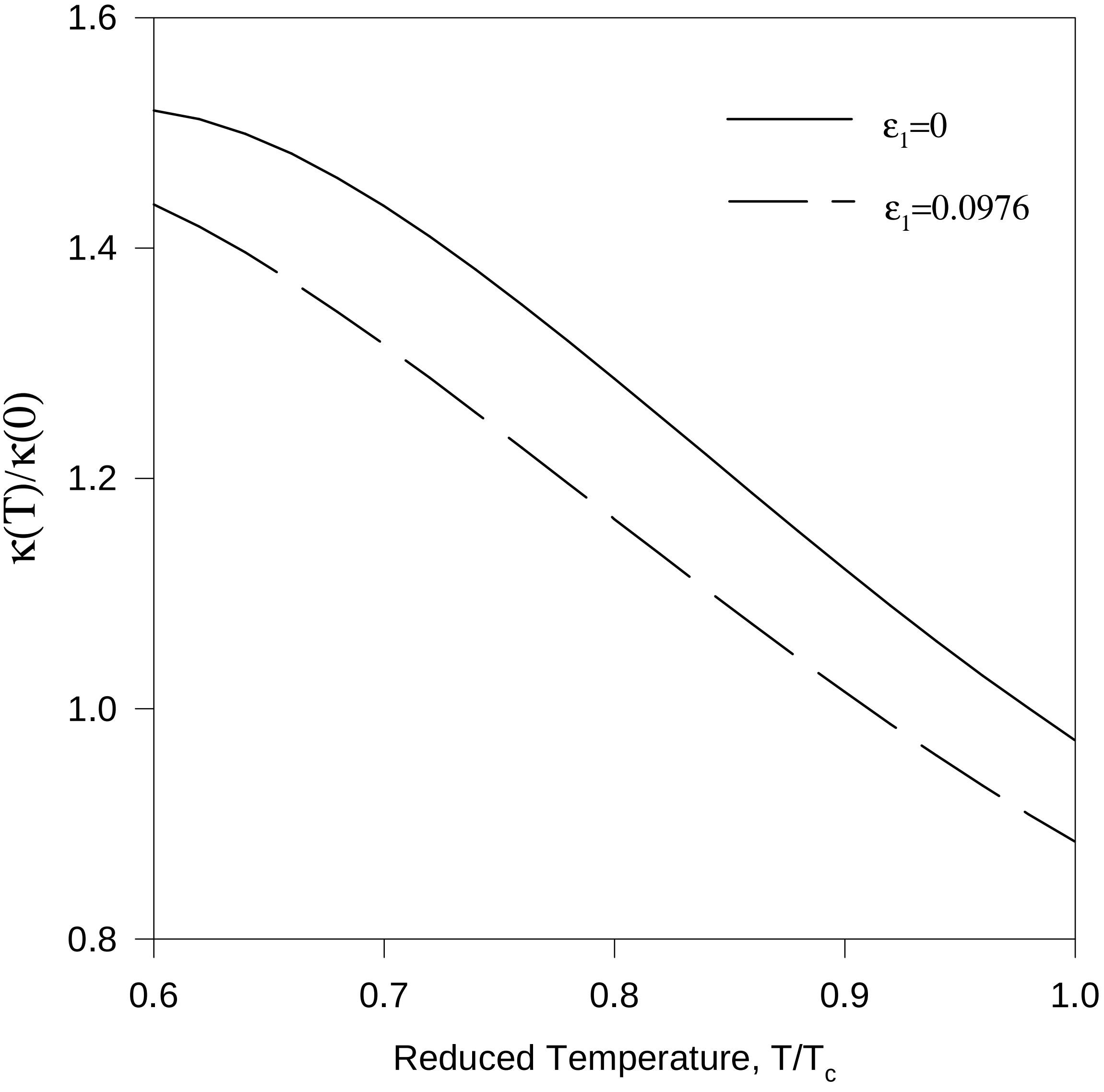

Figure 2